\documentclass[twocolumn,showpacs,preprintnumbers,amsmath,amssymb]{revtex4}

\usepackage{graphicx}
\usepackage{dcolumn}
\usepackage{bm}

\begin{document}

\preprint{ UMD-PP-06-054}

\preprint{UFIFT-HEP-06-15}

\title{Reconciling the CAST and PVLAS Results}
\author{ R.N. Mohapatra$^{1}$ and Salah Nasri$^{2}$\\
$^1$Department of Physics, University of Maryland,
College Park, MD 20742, USA\\
$^2$Department of Physics, University of Florida, Gainesville, FL,
32611} \vskip 0.30cm

  \begin{abstract}
 The PVLAS experiment has recently claimed
evidence for an axion-like particle in the milli-electron-Volt mass
range with a coupling to two photons that appears to be in
contradiction with the negative results of the CAST experiment
searching for solar axions. The simple axion interpretation of these
two experimental results is therefore untenable and it has posed a
challenge for theory. We propose a possible way to reconcile these
two results by postulating the existence of an ultralight
pseudo-scalar particle interacting with two photons and a scalar
boson and the existence of a low scale phase transition in the
theory.
  \end{abstract}

 \maketitle


\section{Introduction}
Two recent experiments, CAST\cite{cast} and PVLAS\cite{pvlas}
searching for an ultralight axion-like pseudoscalar particle
(denoted by $a$) with coupling of the form
\begin{eqnarray}
{\cal L}_I~=~\frac{1}{4M_a}a F\tilde{F}
\end{eqnarray}
have given apparently contradictory results. The CAST\cite{cast}
experiment looked for $a$-particles from the Sun produced in the
reaction $\gamma+\gamma^*\rightarrow a$ where $\gamma^*$ is the
plasmon in the solar plasma and $\gamma$ is a real photon. It found
that in a wide mass range that includes the milli-electron-Volt
(meV) mass , there was no signal for $a$ giving an upper limit on
its two photon coupling that implies $M_a\geq 10^{12}$ GeV. On the
other hand in a laboratory experiment, known as PVLAS\cite{pvlas} a
birefringence effect where the plane of polarization of a photon
beam incident on a static magnetic field is rotated, was observed. A
straightforward interpretation of this appears to be in terms of the
production of a pseudo-scalar axion like particle with mass between
1.7-2 meV with the corresponding coupling scale in Eq. (1) $M_a\sim
2\times 10^5$ GeV. In the context of simple axion
models\cite{peccei}, there appears to be no way to reconcile
these two results\cite{sikivie,raffelt} and several alternative
mechanisms involving new particles  have been
suggested\cite{a1,a2,a3,a4,a5} to resolve this puzzle. This
experimental observation has also stimulated a great deal of
discussion on possible new properties and tests for such
particles\cite{dav,raba}, new 
ways to search for them\cite{ringwald} as well as their
astrophysical implications\cite{rus}.

It is the aim of this brief note to present a new model where the
sub-GeV scale effective Lagrangian consists of
 two light scalars fields $\phi,\sigma$ and a light pseudoscalar $a$
($a$ being the meV particle of interest) and a modified form for the
$a$-photon ($aFF$) interaction. With this, we seem to be able to
reconcile these two experimental results provided we assume the
existence of a low temperature phase transition.

Recall that the main challenge of the axion like interpretation of
the above experiments is to prevent the axion production in the Sun
in $\gamma\gamma^*$ collision without affecting the same in the
laboratory. Our basic idea is to avoid this problem by introducing
an interaction of the form $\phi a F \tilde{F}/M^2$ rather than a
 direct $aF\tilde{F}$ interaction and assuming the field $\phi$
to have the following properties: (i) it has a non-zero vev, which
is induced by a field $\sigma$ such that $<\phi>\sim \kappa
<\sigma>$ and $<\sigma>\sim$ keV where $\kappa$ is a function of the
parameters of the model which we will show later to be of order
$10^5$ giving $<\phi>\sim 0.1$ GeV; (ii) the field $\phi$ has a mass
of order of 20-50 MeV. It then follows that in environments having
temperature below a KeV as in the laboratory, we have the effective
interaction $\frac{1}{M_a}aF\tilde{F}$ (since $<\sigma>\neq 0$ gives
$<\phi>\neq 0$) with an effective coupling $\frac{1}{M_a} =
\frac{<\phi>}{M^2}$ whereas in hot environments $T\gg$ KeV, as in
the core of the Sun, we have $<\sigma>=0$ giving  $<\phi>=0$. This
implies that the effective interaction inside the Sun has the form
$a\phi F\tilde{F}$ type only and not $a F \tilde{F}$. Since $\phi$
mass is far above the solar temperature, this avoids not only $a$
production via the process $\gamma\gamma^*\to a$ but also the
process $\gamma\gamma^*\to a\phi$.
 As we will see below, if $M^2\sim 10^5$ GeV and $m_\phi \sim
20-50$ MeV, the model is consistent with cosmological, astrophysical
as well as laboratory observations, providing a viable way to
reconcile the PVLAS and CAST observations. We also advance a
preliminary speculation on the possible origin of the effective
interaction in terms of new physics.

\section{The scenario and an effective Lagrangian}
 Our model consists of the
neutral spin zero fields, three real scalars $\phi$ and $\sigma$, $S$ and
one real pseudo-scalar $a$, which are singlets under the SM gauge group
and have the effective interaction Lagrangian below  GeV scale given
by
\begin{eqnarray}
{\cal L}~=~ {\cal L}_{kin}+\frac{1}{4M^2}\phi a F \tilde{F}+ 
\lambda_1\phi \sigma
S^2+\lambda_2\sigma S \phi^2\\ \nonumber -\frac{1}{2}m^2_\phi \phi^2
-\frac{1}{2} m^2_a a^2 +\frac{1}{2}M^2_S S^2 -\lambda_S S^4 \\
\nonumber  +\frac{1}{2}m^2_\sigma \sigma^2 -\lambda_\sigma \sigma^4.
\end{eqnarray}
We have displayed only those terms in the Lagrangian which are relevant 
to our discussion and omitted others.
We assume all parameters to be positive and couplings to be of order
one and take as a working example, the masses $M\sim 10^{2.5}$ GeV,
$m_\phi\sim 30$ MeV, $M_S\sim 10^{3}$ GeV, $m_a\sim $ meV and
$m_\sigma\sim $ keV. At zero temperature, this Lagrangian has the
property that $<S>\equiv v_S\sim M_S \sim 10^{3}$ GeV and
$<\sigma>\equiv v_\sigma\sim $ keV. We see from the Lagrangian that
when $<\sigma>\neq 0$ it induces a vev of $\phi$ of order
\begin{eqnarray}
<\phi>\equiv v_\phi\sim
\frac{\lambda_1v_{\sigma}v_S^2}{m^2_\phi+\lambda_2v_{\sigma}v_S},
\end{eqnarray}
We assume that $\lambda_1\sim \frac{v_\phi}{v_S}\sim 10^{-3}$ and
$\lambda_2\sim 1$. We find that for $m^2_\phi\sim \lambda_2v_\sigma
v_S\sim$ $( 0.03 GeV)^2$, $v_\phi\sim 0.5$ GeV.
When the system is in an environment where $T\gg <\sigma>$, we have
phase transition which gives $<\sigma>= 0$ and hence $<\phi>= 0$.

In this framework, we can easily see how to reconcile the PVLAS and
CAST results. In the laboratory, $T\sim 0$ and we have $v_\phi\sim
0.5$ GeV giving $\frac{1}{4M_{a}}a F\tilde{F}$ coupling with $M_{a}
\sim 2\times 10^{5}$ GeV which  explains the PVLAS result. When
temperature of an environment $T\gg $ KeV as in the core of the Sun,
there is phase transition in the $\sigma$ field and we get
$<\sigma>\equiv v_\sigma =0$. This in turn implies that
$<\phi>\equiv v_\phi=0$. Thus in the Sun, $v_\phi= 0$ (since
$v_\sigma =0$) and as a result there is no direct axion-two-photon
interaction. There is of course $\phi a FF$ interaction; but since
$m_\phi\sim 30$ MeV or so, $FFa\phi$ interaction can produce the
$a\phi$ final state only from the ``tail'' of the thermodynamic
distribution of the photons. The cross section is then proportional
to $\frac{T^2}{M^4}e^{-\frac{m_\phi}{T_\odot}}\leq 10^{-10^3}$
GeV$^{-2}$ which is negligible and obviously well below the present
CAST bound for the $a$ production rate. This provides a
reconciliation between the CAST and PVLAS results.

\begin{figure}[tbp]
\begin{center}
\includegraphics[width=2in,height=2in]{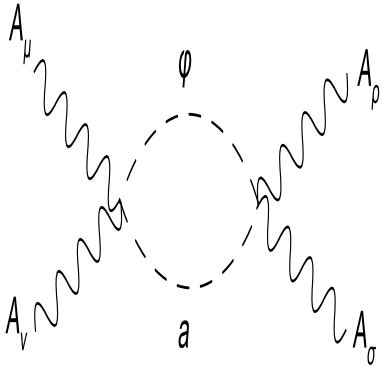}
\caption{Diagram for photon-photon scattering}
\end{center}
\end{figure}

\begin{figure}[tbp]
\begin{center}
\includegraphics[width=2in,height=2in]{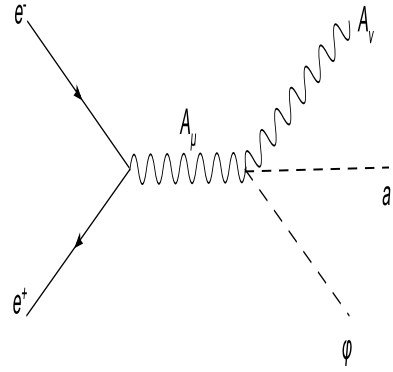}
\caption{The annihilation process $e^+e^- \rightarrow \gamma + a +
\phi$}
\end{center}
\end{figure}

\section{Other implications}
Before considering the phenomenological implications of this
effective Lagrangian model, let us make sure that the theory after
symmetry breaking is consistent i.e. all scalar fields have positive
$m^2$. To check this, let us write down the scalar field mass matrix
for the $\phi,\sigma,S$ system in the basis $(\phi,\sigma, S)$ after
shifting the fields by their vevs.
\begin{eqnarray}
{\cal M}^2= \!\left(
  \begin{array}{ccc}
    \!m^2_{\phi} & \!\lambda_{1}v^2_s + \lambda_2
v_sv_{\phi} & \!\lambda_{1}v_{\sigma}v_s+\lambda_2v_{\sigma}
v_{\phi} \\
     \!\lambda_{1}v^2_s + \lambda_2 v_sv_{\phi} & \! m^{\prime,2}_{\sigma}
     &\!
\lambda_1v_Sv_{\phi} + \lambda_2v_{\phi}^2 \\
     \!\lambda_{1}v_{\sigma}v_s+\lambda_2v_{\sigma} v_{\phi} &
\!\lambda_1v_Sv_{\phi} + \lambda_2v_{\phi}^2  & \!M^{\prime 2}_S
  \end{array}
\right)
\end{eqnarray}
We assume that there is a fine tuning between the two entries in the
$\phi-\sigma$ element of the mass matrix in such a way that the
eigenvalues of this mass matrix are positive. For this to happen, we
must assume that $\frac{\lambda_1}{\lambda_2}\sim
\frac{v_\phi}{v_s}\sim 10^{-3}$. This is the same assumption for the
parameters that we used earlier.

We now make add few comments on the phenomenology of this model:

(i) First point to note is that our interaction leads to light by
light scattering $\gamma\gamma\to \gamma\gamma$ with a strength
$A_{\gamma\gamma\to \gamma\gamma}\sim \frac{1}{M^4}$ arising from a
one loop diagram involving the $\phi -a$ intermediate state in the
loop (see Fig. 1). This is much smaller than the electron
contribution to this process which goes like $A_{\gamma\gamma\to
\gamma\gamma}\sim \frac{\alpha^2}{m^4_e}$ by a factor
$\alpha^{-2}(m_e/M)^4\sim 10^{-18}$. The contribution of this to
$(g-2)$ of muon and electrons depends on the ultraviolet behavior of
the theory. If we assume that above a $100\;$ GeV, the theory is
strongly damped by form factors giving effectively $M\sim 100\;$ GeV
as the cutoff then its contribution to $g-2$ of muons and electrons
is safely consistent with the present experiments.

(ii) In the presence of this interaction, there will be new
contribution to $e^+e^-$ annihilation of type
 $e^+e^-\to \gamma\phi a$ as displayed in
figure $2$. Since $\phi$ has a decay width of approximately
$\Gamma_\phi\sim \frac{m^5_\phi}{\pi^3M^4}\sim 10^{-14}$ GeV (or
lifetime of about $10^{-10}$ sec.), it will decay within a detector
to 2 photons plus an $a$ particle. Thus the signal will look like
$e^+e^-\to 3\gamma+ 2 a$. For $m_{\phi} < GeV$,the cross section for
this process is
\begin{equation}
\sigma(e^+e^- \rightarrow \gamma + a + \phi) \simeq
\frac{\alpha}{(4\pi)^2}\frac{E^2}{M^4}
\end{equation}
 which for $E\sim 100$\; GeV
is of order $10^{-2}$ pico barns  which is smaller than the
predicted standard model background process $e^+e^- \rightarrow
\nu\overline{\nu}+ 3\gamma$ by one order of magnitude \cite{LEP}.

(iii)The new interaction could also lead to the quarkonium (such as
$J/\psi$, $\Upsilon$) decay
\begin{equation}
{\cal{Q}} \rightarrow \gamma + a + \phi
\end{equation}
since we have $m_{\phi}< m_{\cal{Q}} $ . In our model, the branching
ratio for the radiative  decay of $\Upsilon(1S) $ into axion and
$\phi$ is
\begin{equation}
Br(\Upsilon(1S) \rightarrow \gamma + a + \phi) \simeq
2.1.10^{-7}(\frac{100\;GeV}{M})^4
\end{equation}
This is about two order of magnitude below the experimental limit
$Br(\Upsilon(1S) \rightarrow \gamma + \;invisible) < 3.10^{-5}$ and
is safely consistent with our assumptions.

(iv) Second point to note is that this four particle interaction
also does not lead to new channels for rapid energy loss from the
supernova. This is because the $\gamma\gamma\to a\phi$ scattering
has a strength of order $10^{-5}$ GeV$^{-2}$ which is of the same
order as the Fermi constant $G_F$. As a result, the $a$ and the
$\phi$ particles produced in the $\gamma\gamma$ collision will get
trapped in the supernova core and will therefore contribute to the
energy loss from
 supernova by approximately the same amount as a single neutrino species.
This possibility is not ruled out by the SN1987A observations, which
has considerable uncertainty as to how much total energy is emitted
in neutrinos compared to the total energy generated in supernova
core collapse.

(v) To see if this new interaction affects considerations of big
bang nucleosynthesis, we note that above $T\sim $ GeV in the early
universe, the $\phi$ and $a$ were in thermal equilibrium with the
rest of the primordial plasma. As the universe cools below $T\simeq
m_\phi$, the $\phi$ particle decays since its decay rate is bigger
than the Hubble expansion rate at $T\simeq m_\phi$. Thus at the BBN
epoch, in addition to the standard model particles the only two new
particles in equilibrium are $a$ and $\sigma$. Together they
contribute $\delta N_\nu\simeq \frac{8}{7}$. This appears to be
consistent with latest BBN results\cite{olive}.

(vi) Finally, we advance a preliminary speculation on the possible
origin of the higher dimensional $a\phi F\tilde{F}$ interaction.
Consider extending the standard model by adding a vector like pair
of fermions $\psi_{L,R}$ which are leptonic and have electric charge
$\pm 1$ (or equivalently $Y=\pm 2$ and $M_{\psi}\geq $ 100 GeV.  Let
there be a neutral pseudo-scalar particle $A$ with mass $M_A\sim$
few GeV with an interaction of the form
\begin{equation}
{\cal L}_I~=~f A\bar{\psi}\gamma_5 \psi +M_0 A a\phi~+~h.c
\end{equation}
 In this model the familiar triangle diagram will induce an effective
$AF\tilde{F}$ interaction with approximate strength
$\frac{\alpha}{4\pi M_\psi}$. The tree level exchange of $A$ will
then generate an effective Lagrangian below the GeV scale of the
form we have been discussing. The parameters of the theory can be so
chosen as to give $M_a\sim 10^5 $ GeV (e.g $M_0\sim$ GeV).  This
would provide a renormalizable version of our model. Of course one
could perhaps interpret $A$ as an axion corresponding to a different
strongly interacting theory (technicolor ?). Further discussion of
this model as well as its detailed implications is postponed to a
future publication.

 In conclusion, we have presented an effective
Lagrangian below the GeV scale which seems to be able to reconcile
the CAST and the PVLAS experiments in a way that is different from
other existing explanations. We find the model to be 
compatible with all laboratory, astrophysical as well as
cosmological observations. We also advance a preliminary speculation
on a possible origin of this effective Lagrangian. Detailed
phenomenological and astrophysical implications of this model are
currently under investigation.

We like to thank P. Sikivie for useful discussions and
encouragement. We also thank T. Okui for some comments.
 The work of R.N.M is supported by the National Science Foundation grant
no. Phy-0354401. S.N is supported by the DOE grant No. DE-FG02-97ER41029.

\end{document}